# Classical Machine Learning Baselines for Deepfake Audio Detection on the Fake-or-Real Dataset

Faheem Ahmad, Ajan Ahmed, Masudul Imtiaz
*Department of Electrical and Computer Engineering*
*Clarkson University*
Potsdam, New York, USA
fahmad@clarkson.edu, aahmed@clarkson.edu, mimtiaz@clarkson.edu

*Abstract*—Deep learning has enabled highly realistic synthetic speech, raising concerns about fraud, impersonation, and disinformation. Despite rapid progress in neural detectors, transparent baselines are needed to reveal which acoustic cues reliably separate real from synthetic speech. This paper presents an interpretable classical machine learning baseline for deepfake audio detection using the Fake-or-Real (FoR) dataset. We extract prosodic, voice-quality, and spectral features from two-second clips at 44.1 kHz (high-fidelity) and 16 kHz (telephone-quality) sampling rates. Statistical analysis (ANOVA, correlation heatmaps) identifies features that differ significantly between real and fake speech. We then train multiple classifiers—Logistic Regression, LDA, QDA, Gaussian Naïve Bayes, SVMs, and GMMs—and evaluate performance using accuracy, ROC–AUC, EER, and DET curves. Pairwise McNemar's tests confirm statistically significant differences between models. The best model, an RBF SVM, achieves ∼93% test accuracy and ∼7% EER on both sampling rates, while linear models reach ∼75% accuracy. Feature analysis reveals that pitch variability and spectral richness (spectral centroid, bandwidth) are key discriminative cues. These results provide a strong, interpretable baseline for future deepfake audio detectors.

## I. INTRODUCTION

Neural text-to-speech (TTS) and voice conversion systems now generate highly realistic synthetic speech from text or brief voice samples [1], [2]. While beneficial for accessibility and entertainment, these technologies enable malicious applications including voice impersonation, fabricated audio evidence, and automated social engineering [3], posing significant security threats.

Current deepfake audio detection methods [3]–[5] predominantly employ deep neural networks operating on spectrograms, waveforms, or speaker embeddings. Despite achieving strong performance, these models require substantial computational resources and lack interpretability, obscuring the acoustic cues they exploit for discrimination.

Classical machine learning with hand-engineered features offers complementary advantages: computational efficiency suitable for resource-constrained devices (telephony gateways, browser extensions) and interpretability enabling analysis of acoustic differences between genuine and synthetic speech. This interpretability can inform understanding of TTS evolution and potential evasion strategies.

This paper establishes a comprehensive, interpretable baseline for deepfake detection on the Fake-or-Real (FoR) dataset [6], emphasizing: (i) acoustic features capturing prosody, voice quality, and spectral structure; (ii) statistical analysis (ANOVA, correlation) for feature selection; (iii) classical models including Logistic Regression, LDA, QDA, Naïve Bayes, SVMs, and GMMs; and (iv) robust evaluation with accuracy, ROC–AUC, EER, DET curves, and McNemar's tests.

We evaluate two conditions: high-quality 44.1 kHz audio and 16 kHz re-recorded audio with room playback, simulating realistic channel degradation. Sections II–V detail dataset characteristics, feature engineering, modeling methodology, and experimental results. Section VI discusses implications, and Section VII presents conclusions and future directions.

## II. FAKE-OR-REAL DATASET

The Fake-or-Real (FoR) dataset [6] is a publicly available corpus for synthetic speech detection containing short audio clips of authentic human speech from corpora including CMU Arctic [7], LJSpeech [8], and VoxForge [9], alongside synthetic speech generated by modern TTS systems such as WaveNet [1]. This diversity enables robust evaluation of deepfake detection algorithms [10]. The dataset includes four variants: **for-original** (varying formats), **for-norm** (standardized preprocessing), **for-2sec** (2-second segments), and **for-rerec** (room-recorded with telephone-like degradation). We utilize the **for-2sec** (44.1 kHz) and **for-rerec** (16 kHz) variants, which provide speaker-disjoint training and test partitions to prevent speaker-specific memorization. The dataset comprises approximately 31,138 clips with balanced real/fake distribution (∼50% each): 24,913 for training and 6,225 for testing. Given this natural balance, we avoid oversampling or SMOTE, which could introduce artificial patterns in the feature space.

## III. FEATURE EXTRACTION AND SELECTION

For each two-second clip we compute frame-level measurements of acoustic features and then summarize them with statistics such as mean, standard deviation, range, percentiles, and coefficients of variation.

### A. Prosodic Features

Prosodic features describe the perceived melody and rhythm of speech. We compute the fundamental frequency ($F_0$) using

the YIN algorithm [11] and derive the following statistics over voiced frames:
- *f0_mean_v*: Mean $F_0$ in Hz during voiced segments.
- *f0_std_v*: Standard deviation of $F_0$, indicating pitch variability.
- *f0_range_v*: Range (max–min) of $F_0$.
- *f0_iqr_v*: Interquartile range (IQR) of $F_0$.
- *f0_cv_v*: Coefficient of variation (*f0_std_v/f0_mean_v*).
- *f0_p10_v*, *f0_p90_v*: 10th and 90th percentile of $F_0$.

We also extract timing-related prosodic features based on voiced/unvoiced segmentation:
- *dur_s*: Duration of the trimmed clip in seconds.
- *voice_pct*: Proportion of time that is voiced.
- *n_voiced_seg_per_s*: Number of distinct voiced segments per second.
- *mean_voiced_seg_ms*: Average voiced segment duration in ms.
- *pause_ratio*: Ratio of unvoiced (pause) duration to voiced duration.
- *f0_slope_hz_per_s*: Slope of a linear regression fit to the $F_0$ contour.

These features are motivated by the observation that human speech often exhibits rich, context-dependent prosody, whereas synthetic speech may have flatter intonation or overly regular timing. Many TTS systems historically struggled to mimic natural pitch variability, resulting in more monotone voices.

### B. Voice-Quality Features

Voice quality describes fine-grained characteristics of vocal fold vibration and the resulting glottal source [12], [13]. We include:
- *jitter_local*: Relative cycle-to-cycle variation in pitch period.
- *shimmer_local*: Relative cycle-to-cycle variation in amplitude.

In natural speech, jitter and shimmer arise from small irregularities in vocal fold vibration [14]. Synthetic speech generated from vocoders can be overly smooth, with very low jitter and shimmer. Thus, these measures can indicate how "perfect" or "natural" the glottal source appears.

### C. Energy and Spectral Features

We extract spectral descriptors from the STFT magnitude spectrum (25 ms window, 10 ms hop), computing statistics (mean, std, range, IQR, CV where applicable) for each:

**RMS Energy:** Frame-level energy statistics capture loudness and signal dynamics. Synthetic speech typically exhibits more uniform amplitude normalization than real recordings.

**Spectral Centroid:** The frequency "center of mass" correlates with perceived brightness. Real speech often shows higher centroids due to sibilants, plosives, and microphone noise absent in synthetic voices.

**Spectral Bandwidth:** Measures energy spread around the centroid. Natural speech typically exhibits larger bandwidths from broadband noise and diverse formant structures.

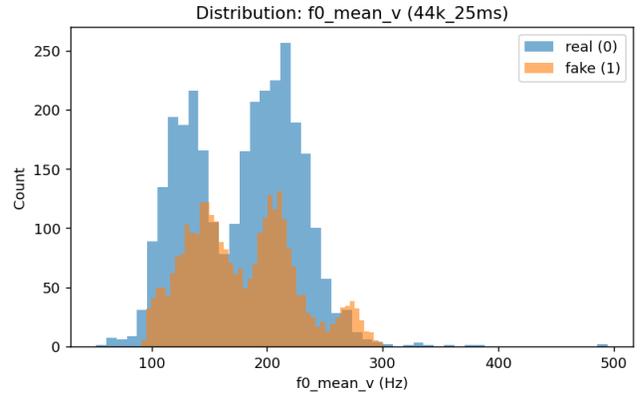

Fig. 1. Distribution of mean fundamental frequency (*f0_mean_v*) for real (blue) and fake (orange) speech at 44.1 kHz. Real speech exhibits two broad peaks corresponding to male and female pitch ranges; fake speech covers a similar range but with a slightly different distribution.

**Spectral Contrast and Rolloff:** Spectral contrast quantifies peak-valley differences across frequency bands, while rolloff identifies the frequency containing 85% of spectral energy. These features characterize formant sharpness and high-frequency energy distribution.

### D. Feature Selection via ANOVA

One-way ANOVA on the training set identified highly discriminative features ($p \approx 0$): *f0_std_v*, *f0_cv_v*, *f0_range_v*, *f0_iqr_v*, *rms_mean*, *spec_centroid_mean*, *spec_bandwidth_mean*, and *spec_rolloff_mean*. These results confirm that pitch variability and spectral richness effectively distinguish natural from synthetic speech. Features with $p \geq 0.05$ (e.g., *shimmer_local* with $p \approx 0.91$) were excluded, yielding 29–30 features per condition.

### E. Feature Visualizations

Figures 1–3 visualize feature distributions at 44.1 kHz. Real speech shows bimodal $F_0$ distributions reflecting gender differences (Fig. 1), higher spectral centroids from sibilants and microphone noise (Fig. 2), and occupies high-centroid, high-bandwidth regions rarely reached by synthetic speech (Fig. 3). The 16 kHz condition (Figs. 4, 5) shows similar patterns, confirming that discriminative cues survive downsampling and re-recording. Pitch distributions remain bimodal (Fig. 4), and real speech maintains higher centroid-bandwidth characteristics (Fig. 5). Correlation heatmaps (Fig. 6) reveal feature interdependencies. Spectral features (centroid, bandwidth, rolloff) and energy features form distinct correlation blocks. Importantly, differences between real and fake heatmaps indicate shifts in feature relationships, providing additional discriminative power.

## IV. MODELING PIPELINE AND EVALUATION

### A. Data Splitting and Preprocessing

For each sampling-rate condition, we use the dataset's predefined training and test sets. Speaker identities do not

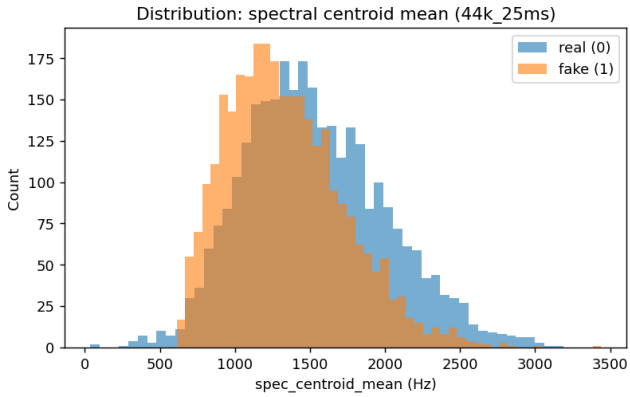

Fig. 2. Distribution of mean spectral centroid at 44.1 kHz. Real speech tends to have a higher centroid, indicating more high-frequency energy (e.g., sibilants and microphone noise) than synthetic speech.

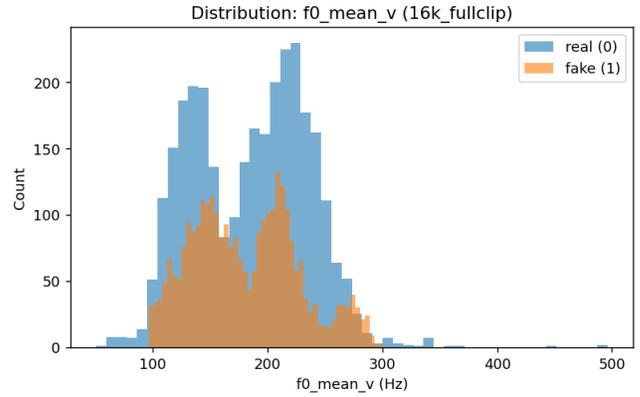

Fig. 4. Distribution of $f0\_mean\_v$ for real and fake speech at 16 kHz. The overall structure resembles the 44.1 kHz case, indicating that downsampling and re-recording do not remove pitch-based cues.

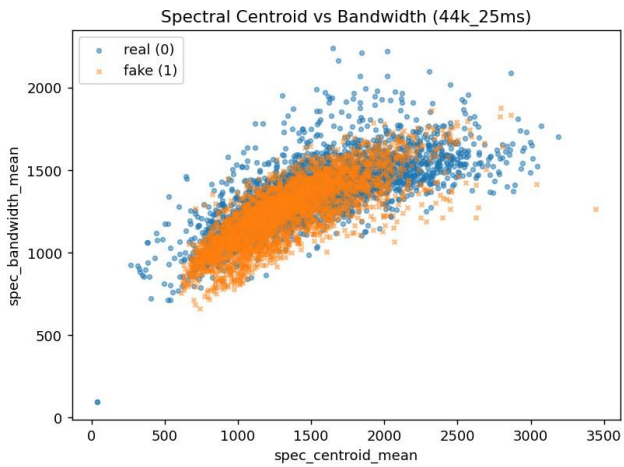

Fig. 3. Scatter plot of spectral centroid versus bandwidth at 44.1 kHz. Real clips (blue) and fake clips (orange) overlap, but many real clips occupy the high-centroid, high-bandwidth region that fake clips rarely reach.

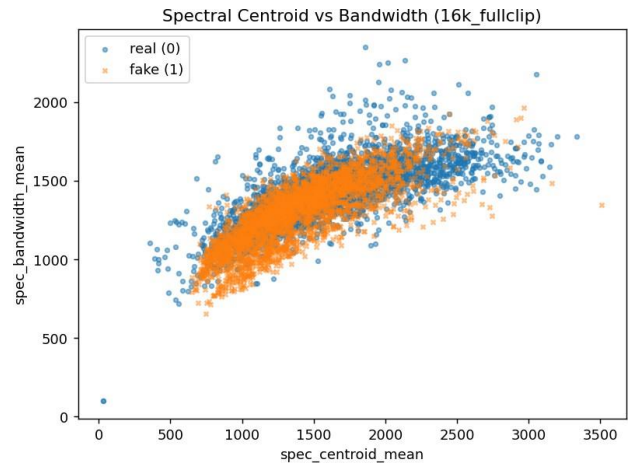

Fig. 5. Spectral centroid versus bandwidth at 16 kHz. Real speech still tends to occupy regions with higher centroid and bandwidth, reflecting broader spectral content even after re-recording.

overlap between splits, preventing speaker leakage. We apply the following preprocessing steps:

1) **Feature cleaning:** Replace any occurrences of $\pm\infty$ with NaN and drop features that are entirely missing.
2) **Imputation:** For remaining missing values (rare in practice), we use median imputation based on the training set.
3) **Scaling:** All features are standardized to zero mean and unit variance using statistics computed on the training set. The same transformation is applied to the test set.

### B. Classical Machine Learning Models

We evaluate the following models:

- **Logistic Regression (LR):** A linear model with L2 regularization and class weights to handle minor imbalance [15].
- **Linear Discriminant Analysis (LDA):** Assumes each class is Gaussian with shared covariance, yielding a linear decision boundary [16].
- **Quadratic Discriminant Analysis (QDA):** Similar to LDA but with class-specific covariance matrices, allowing quadratic boundaries.
- **Gaussian Naïve Bayes (GNB):** Models each feature as an independent Gaussian given the class. This is simple but often underfits when features are correlated.
- **Support Vector Machines (SVM):** We use a linear SVM and an SVM with a radial basis function (RBF) kernel [17]. For the RBF SVM, we perform grid search over $C$ and $\gamma$, using stratified 3-fold cross-validation. The best configuration was $C = 10$ and $\gamma$ = scale.
- **Gaussian Mixture Model (GMM) classifier:** We train separate GMMs for the real and fake classes and classify by comparing the log-likelihoods [18].

### C. Evaluation Metrics

We evaluate both training and test performance using:

- **Accuracy:** Fraction of correctly classified clips.

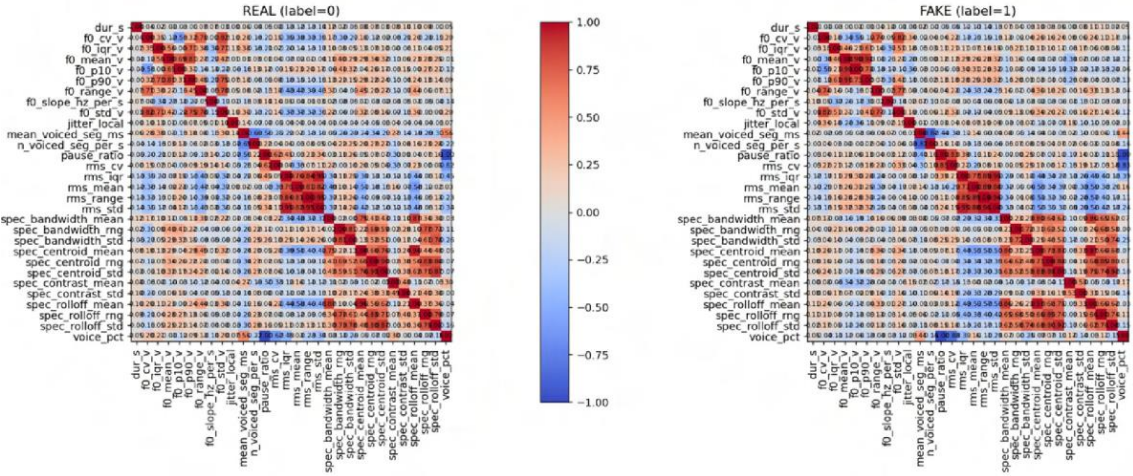

Fig. 6. Feature correlation heatmaps for real (left) and fake (right) speech at 16 kHz. Blocks of high correlation appear among spectral features (centroid, bandwidth, rolloff) and among energy features. Differences between the two heatmaps indicate how feature relationships shift between genuine and synthetic speech, providing additional discriminative information.

- **ROC–AUC:** Area under the receiver operating characteristic curve [19].
- **Equal Error Rate (EER):** Error rate where FAR equals FRR [20].
- **DET curves:** Plots of FAR versus FRR across thresholds.

To assess whether differences between models are statistically significant, we compute pairwise McNemar's tests [21], [22].

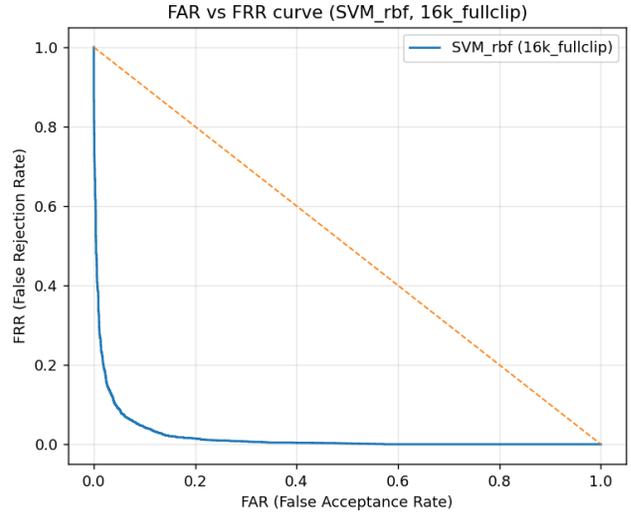

Fig. 8. DET curve for RBF SVM on 16 kHz condition. The equal error rate is slightly lower, around 6.6%.

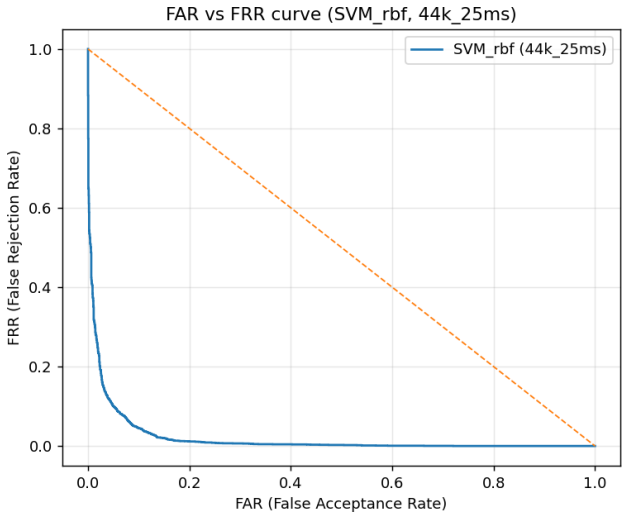

Fig. 7. DET curve for RBF SVM on 44.1 kHz condition. The EER point occurs where FAR and FRR intersect at approximately 7.3%.

## V. RESULTS

### A. Model Performance at 44.1 kHz

Table I summarizes the performance of all models on the 44.1 kHz (for-2sec) condition. We report training and test accuracy, ROC–AUC, and test EER. The linear models (Logistic Regression, LDA, Linear SVM) cluster around 75% accuracy and AUC ≈ 0.82. QDA and GMM improve to approximately 77–78% accuracy and AUC ≈ 0.85–0.87. The RBF SVM substantially outperforms all others, achieving nearly 93% test accuracy and an EER of only 7.3%.

### B. Model Performance at 16 kHz

Table II shows analogous results for the 16 kHz (for-rerec) condition. Interestingly, performance does not degrade in the re-recorded, lower-bandwidth condition; in fact, most models slightly improve. The RBF SVM reaches an EER of 6.6%, while GMM and QDA also see modest gains. This indicates that the channel effects introduced in the re-recorded data do not remove the discriminative cues; they may even introduce additional artifacts that differ between real and fake clips.

TABLE I
MODEL PERFORMANCE ON FOR (44.1 KHZ, 2-SECOND CLIPS).

| Model | Train | | Test | | |
|---|---|---|---|---|---|
| | Acc | AUC | Acc | AUC | EER |
| Logistic Reg. | .751 | .823 | .753 | .828 | 24.7 |
| LDA | .747 | .816 | .753 | .819 | 24.7 |
| QDA | .768 | .854 | .771 | .854 | 22.9 |
| Naïve Bayes | .698 | .768 | .696 | .764 | 30.4 |
| Linear SVM | .754 | .823 | .755 | .826 | 24.5 |
| **RBF SVM** | **.953** | **.990** | **.927** | **.980** | **7.3** |
| GMM | .787 | .875 | .783 | .869 | 21.7 |

TABLE II
MODEL PERFORMANCE ON FOR (16 KHZ, RE-RECORDED CLIPS).

| Model | Train | | Test | | |
|---|---|---|---|---|---|
| | Acc | AUC | Acc | AUC | EER |
| Logistic Reg. | .756 | .831 | .754 | .834 | 24.6 |
| LDA | .753 | .826 | .752 | .827 | 24.8 |
| QDA | .774 | .858 | .771 | .857 | 22.9 |
| Naïve Bayes | .701 | .772 | .698 | .769 | 30.2 |
| Linear SVM | .758 | .829 | .758 | .830 | 24.2 |
| **RBF SVM** | **.957** | **.991** | **.934** | **.981** | **6.6** |
| GMM | .811 | .895 | .805 | .888 | 19.5 |

### C. ROC and DET Curves

Fig. 9 shows ROC and DET curves for all models on the 44.1 kHz condition. The ROC plot confirms that the RBF SVM dominates with AUC close to 0.98, while the DET plot shows its operating points lie well below other models across a wide range of false positive rates. Linear models cluster together with similar performance, while QDA and GMM provide moderate gains.

We also present individual DET curves for the RBF SVM to highlight the EER points. Fig. 7 shows the DET curve for the 44.1 kHz condition, where the EER occurs at approximately 7.3%. Fig. 8 shows the corresponding curve for the 16 kHz condition, where the EER is slightly lower at 6.6%. These curves demonstrate that one can operate at FAR and FRR around 5% simultaneously, or drive FAR even lower at the cost of somewhat higher FRR, depending on application requirements.

### D. Statistical Significance (McNemar's Tests)

To establish whether differences between models are statistically significant, we computed McNemar's tests on pairwise model disagreements. Key findings include:

- Logistic Regression, LDA, and Linear SVM show no significant pairwise differences on the test set ($p > 0.4$). Their small accuracy differences are not meaningful.
- QDA and GMM are significantly better than linear models ($p \ll 0.05$), confirming non-linear boundaries improve detection.
- The RBF SVM is significantly better than every other model ($p \approx 0$), supporting the conclusion that its performance advantage is not due to random variation.
- Naïve Bayes is significantly worse than all other models, consistent with its oversimplified feature-independence assumptions.

This confirms the ranking: RBF SVM $\gg$ (GMM $\approx$ QDA) $>$ (LR $\approx$ LDA $\approx$ Linear SVM) $\gg$ GNB.

## VI. DISCUSSION

### A. What Makes Speech Sound "Real"?

The ANOVA and visualization results point to several key differences between real and synthetic speech. **Pitch variability:** Features such as *f0_std_v*, *f0_range_v*, and *f0_cv_v* were among the most significant. Real speech displays richer, context-dependent intonation, whereas synthetic speech often exhibits flatter pitch contours. Even if the average pitch is realistic, the fine dynamics of rising and falling intonation are harder to synthesize. **Spectral richness:** Real recordings typically have greater high-frequency content and broader spectral bandwidth, manifesting in higher values for *spec_centroid_mean*, *spec_bandwidth_mean*, and *spec_rolloff_mean*. Synthetic speech, especially vocoder-based, may sound slightly muffled, lacking sharp consonant bursts and recording noise. **Voice quality:** Jitter shows meaningful differences, suggesting synthetic voices are often too "perfect" in their periodicity. Shimmer was not discriminative, indicating amplitude irregularities are similar between classes.

### B. Why Classical Models Perform Well

The RBF SVM's strong performance (AUC $\approx$ 0.98, EER $\approx$ 7%) indicates that real and fake speech are nearly separable in the handcrafted feature space after a modest non-linear transformation. Classical models provide interpretability, data efficiency, and computational efficiency suitable for real-time deployment on modest hardware.

### C. Limitations and Generalization

The FoR dataset uses specific TTS engines; the RBF SVM may exploit dataset-specific artifacts that may not generalize to future synthesis methods. Linear models, though less accurate, may capture more fundamental differences likely to persist across TTS systems. A robust system might combine classical features with deeper representations.

## VII. CONCLUSION

This paper presented a classical machine learning baseline for deepfake audio detection on the Fake-or-Real dataset using prosodic, voice-quality, and spectral features. As shown in Table I and Table II, the RBF SVM achieved 93% test accuracy and EER near 7% on both 44.1 kHz and 16 kHz conditions, significantly outperforming linear models (~75% accuracy). Pitch variability and spectral richness emerged as the most discriminative cues, as illustrated in Fig. 1–5. The correlation analysis in Fig. 6 revealed that feature relationships differ between real and fake speech. Performance did not degrade after re-recording and downsampling, indicating discriminative cues persist under realistic channel distortions. Future work includes adding cepstral features (MFCCs [23], CQCCs [24]), testing cross-corpus generalization on ASVspoof [20] and FakeAVCeleb [25], and exploring hybrid classical-deep systems.

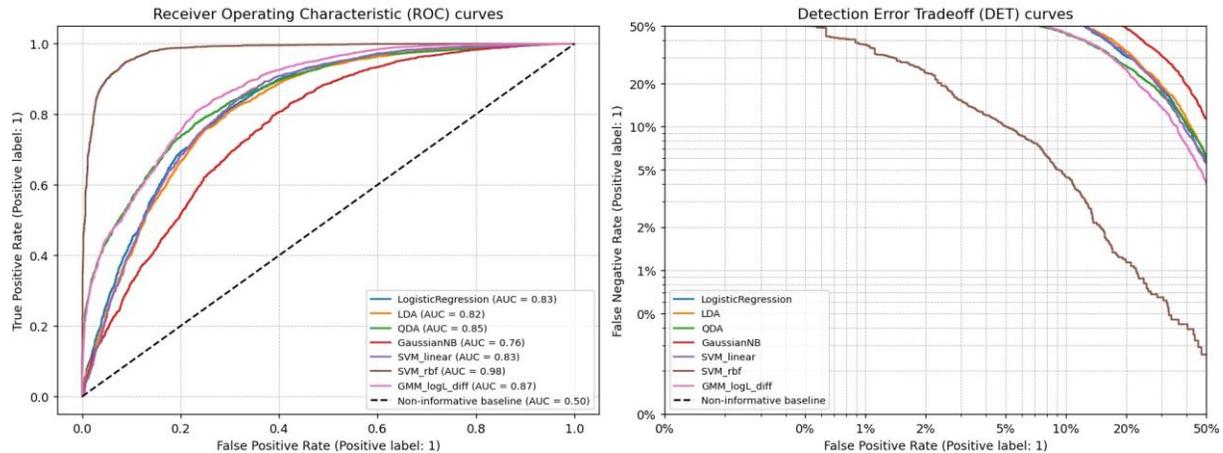

Fig. 9. Receiver Operating Characteristic (ROC) and Detection Error Trade-off (DET) curves for all classical models on the 44.1 kHz FoR condition. Each colored line corresponds to one model. The ROC plot (left) shows the RBF SVM dominates with AUC close to 0.98, while the DET plot (right) shows its operating points lie well below other models across a wide range of false positive rates.